\documentclass[%
 reprint,
 amsmath,amssymb,
 aps
floatfix,twocolumn
]{revtex4}

\usepackage{epsfig}
\usepackage{amsmath,amsfonts,amssymb}
\usepackage{graphicx}
\usepackage{dcolumn}
\usepackage{bm}

\begin{document}

\preprint{APS/123-QED}

\title{Why can hadronic stars convert into strange  quark stars with larger radii}

\author{Alessandro Drago}
\author{Giuseppe Pagliara}
\affiliation{Dipartimento di Fisica e Scienze della Terra, Universit\'a di Ferrara, Via Saragat, 1, 44122 Ferrara, Italy\\
INFN - Sezione di Ferrara, Via Saragat, 1, 44122 Ferrara, Italy}%

\date{\today}

\begin{abstract}
The total binding energy of compact stars is the sum of the gravitational binding energy $(BE)_g$ and the nuclear binding energy $(BE)_n$, the last being related to the microphysics of the interactions. 
While the first is positive (binding) both for hadronic stars and for strange quark stars, the second is large and negative for hadronic stars (anti-binding) and either small and negative (anti-binding) or positive (binding) for strange quark stars.
A hadronic star can convert into a strange quark star with a larger radius because the consequent reduction of $(BE)_g$ is over-compensated by the large increase in $(BE)_n$. Thus, the total binding energy increases due to the conversion and the process is exothermic. Depending on the equations of state of hadronic matter and quark matter and on the baryonic mass of the star, the contrary is obviously also possible, namely the conversion of hadronic stars into strange quark stars having smaller radii, a situation more often discussed in the literature.
We provide a condition that is sufficient and in most of the phenomenologically relevant cases also necessary in order to form strange quark stars with larger radii while satisfying the exothermicity request. Finally, we compare the two schemes in which quark stars are produced (one having large quark stars and the other having small quark stars) among themselves and with the third-family scenario and we discuss how present and future data can discriminate among them.
\end{abstract}

\maketitle

\section{Introduction}
    After the discovery of compact stars having a mass of at least about $2M_{\odot}$ \cite{Antoniadis:2013pzd,Cromartie:2019kug} one problem has emerged: how to reconcile the request of an Equation of State (EoS) stiff enough to support those massive stellar objects with the possible indication, coming from the analysis of a variety of x-ray emissions \cite{Bogdanov:2016nle,dEtivaux:2019cnf} of small radii for at least some of those objects? If confirmed, the existence of compact stars having a mass of about $(1.4-1.5) M_{\odot}$ and radii smaller than about $11.5$ km would imply that the EoS is very soft in a certain range of densities and very stiff in another, signalling the appearance of a strong phase transition in the core of at least the more massive compact stars. In fact, it has been shown that in the absence of such a transition the radius of a $1.4 M_{\odot}$ star  cannot be much smaller than about $12$ km if all the information coming from nuclear physics and from astrophysics are taken into account \cite{Most:2018hfd}.

One possible solution of this problem is to assume that the quark deconfinement phase transition produces a hybrid star and that the transition is strong enough that a "third-family" of objects is generated \footnote{The first family would be that of white dwarfs and the second that of neutron stars containing no more than a small amount of quarks.}. It has been shown though that in that scenario, in order to obtain radii smaller than about $11.5$ km the speed of sound of quark matter has to be close to the speed of light \cite{Alford:2015dpa}, a requests that certainly is not easy to satisfy with microscopically motivated quark EoSs.

In Ref.~\cite{Drago:2013fsa} a different solution has been proposed, based on the idea of the coexistence of two families of stars: hadronic stars (HSs) and strange quark stars (QSs). Massive stars (with mass $M\gtrsim 1.6 M_{\odot}$) are interpreted as QSs and can reach large radii, up to about $13$km depending on the adopted model for the quark matter EoS, while less massive compact stars are interpreted as HSs i.e. as stars composed of nucleons, hyperons and delta resonances and they can have small radii, typically $R_{1.4} \lesssim 11$km for the $1.4M_{\odot}$ configuration. During the last years we have developed in great detail that scenario and in Refs.\cite{Drago:2015dea,Drago:2015cea,Drago:2015qwa,Dondi:2016yjl,Pili:2016hqo,Wiktorowicz:2017swq,Drago:2017bnf,Burgio:2018yix,Drago:2019tbs,DePietri:2019khb,Char:2019wvo} one can find discussions of its rich phenomenology, including its impact on the merger of two compact stars.
Presently, this scenario remains a viable possibility and actually some recent findings seem to support it: for instance in Ref.\cite{Sarin:2020pwr}, by considering the light curves of a sample of X-ray afterglows associated with short gamma-ray bursts,
it has been found that the objects formed immediately after the merger are not well described by a hadronic equation of state, although at $1\sigma$ level only. That analysis seems therefore to indicate that during the merger, quark deconfinement takes place and that the outcome of the merger is either a Black Hole or a QS, as suggested by the two-families scenario \cite{Drago:2015qwa,Drago:2017bnf,DePietri:2019khb}.

The idea of coexistence of two families of compact stars
dates back to Refs.\cite{Berezhiani:2002ks,Bombaci:2004mt,Drago:2004vu} where the conversion of a HS into a QS was proposed as the 
source of some long gamma-ray bursts lacking a clear supernova counterpart. The link between this conversion process and the inner engine 
of gamma-ray-bursts had already been proposed in \cite{Bombaci:2000cv}.
In those papers, for what concerns the structure of compact stars, it was generally assumed that at a fixed value of the baryonic mass $M
_b$ (which is defined as the total number of baryons multiplied, conventionally,  by the atomic mass unit) QSs have a smaller radius with respect to their progenitor HSs, i.e. the burning of hadronic matter into quark matter is accompanied by a fast contraction of the radius of the star. Notice though that already in Ref.\cite{Drago:2005yj} a counterexample was found in which hybrid stars (with unpaired quarks) can have a smaller radius and a larger gravitational mass with respect to a pure color superconducting QS with the same $M_b$. Thus the formation of a color superconducting condensate can lead to the conversion of the initial hybrid star configuration into a pure QS configuration with a larger radius \footnote{see Fig.9 of the above mentioned paper.}.
This means that an exothermic process, such as the conversion of a HS into a QS, is not necessarily accompanied by a "compactification" of the star, although this possibility was not discussed in an explicit and detailed way in the literature. 

After the discovery of very massive compact stars ($M \gtrsim 2M_{\odot}$ ), the scenario of coexistence of HSs and QSs has therefore been reformulated in Ref.~\cite{Drago:2013fsa} with the use of quark matter EoSs that are stiff enough to fulfil the constraint on the maximum mass and with the use of hadronic matter EoSs which are soft enough to justify small values for $R_{1.4}$. Therefore, at least in principle, the combination of stiff quark matter EoSs and of soft hadronic matter EoSs can produce QSs, originated by the conversion of a HS, which are larger than their progenitor.
Let us remind that the criterion for having an exothermic process is that at fixed $M_b$, $M_g^Q<M_g^H$, i.e. the QS is lighter than the HS; the radii of the two configurations do not play any explicit role.
This fact is somehow counter-intuitive: one would naively expect a larger configuration to be less bound by gravity and therefore not energetically favored. This is true for
not-self bound stars, i.e. HSs and hybrid stars.
On the other hand, QSs are self-bound objects:
namely their EoS has a 
zero of the pressure at a finite density at which the energy per baryon is smaller
than the one of Iron \cite{Haensel:2007yy}.
In turn, this implies
that the binding energy associated with the
microphysics of the interaction can be quite
relevant in determining if the conversion process
to quark matter, in the presence of gravity, is
exothermic or endothermic.

Let us briefly remark that while QSs are self-
bound objects they still can have a crust, for instance the
thin nuclear crust, similar to the outer crust of HSs,
discussed in Ref.\cite{Haensel:1986qb}. In that case the bulk of the object is self-bound but the thin crust is not. In
Ref.\cite{Drago:2001nq} a self-bound
crust composed of a hadron-quark mixed
phase was suggested, while a crust composed by
quark nuggets embedded in a uniform electron
background was discussed in Ref.\cite{Jaikumar:2005ne}.

In this work we will present a clear argument showing that the conversion of a HS into a QS, with or without a crust, can lead to an increase of the radius while being, at the same time, energetically favored. 

\section{Binding energies of compact stars and exothermic processes}
\subsection{Decomposition of the binding energy}
The binding energy of a star $BE$ is the difference between the baryonic mass and the gravitational mass: $BE=M_b-M_g$ (a positive binding energy corresponds therefore to bound systems).
Let us start by considering a transition between two stars having a crust, for instance a first star composed of hadrons and a second one containing quarks only in its central region, i.e. a hybrid star.
When considering the conversion of a HS into a hybrid star, the energy released $E$ is the difference between the binding energy of the final configuration and that of the progenitor: $E=BE^{hyb}-BE^H$ (at fixed $M_b$). If $E>0$ the process is exothermic.
For stars with a crust (not self-bound stars), one can find in Ref.\cite{Lattimer:2000nx} a simple universal relation 
between $BE$ and the compactness $c=M/R$, that has been tested for a wide collection of EoSs: $BE/M_g=0.6\frac{c}{(1-c/2)}$, see also the more recent and accurate relations in
Ref.\cite{Breu:2016ufb}. 
Let us consider the leading order in $c$: $BE=0.6M_g^2/R$. For a fixed $M_g^H$ and $R^H$ (mass and radius of the HS), $M_b=M_g^H+0.6(M_g^H)^2/R^H=M_g^{hyb}+0.6(M_g^{hyb})^2/R^{hyb}$. By imposing
the condition of exothermicity, $M_g^H-M_g^{hyb}>0$, one easily obtains the necessary condition $R^{hyb}<R^H$.  Thus one can conclude that an exothermic process is necessarily accompanied by a reduction of the radius, i.e. the newly formed hybrid star has a radius smaller than the one of its hadronic progenitor. This is the result obtained in all the calculations in which a transition between two compact stars having a crust is considered and it is presumably at the origin of the idea that an exothermic process taking place inside a compact star always produces a more compact object.

However, this argument in not valid in general. First, the above mentioned universal relation between $BE$ and $c$ does not apply to QSs, as found recently in \cite{Jiang:2019pzw}. A universal relation can however be established
when considering that the binding energy of a star is the sum of two contributions: a gravitational binding energy $(BE)_g$ and a nuclear binding energy $(BE)_n$ which are interpreted as the binding energy obtained when switching off the nuclear and  the gravitational interactions, respectively.
To define these quantities one needs to introduce the so-called proper mass $M_p$:
\begin{equation}
M_p=\int_0^R\mathrm{d}r\,\, 4\pi r^2 \frac{e(r)}{\sqrt{1-2M_g(r)/r}}
\end{equation}
where $e(r)$ is the energy density profile.
As done in Ref.\cite{Bombaci:2000cv}, it is possible to split $BE$ in the following way: $BE=M_b-M_g=(M_b-M_p)+(M_p-M_g)\equiv(BE)_n+(BE)_g$ with the two terms representing the nuclear binding energy and the gravitational binding energy, respectively.
The crucial point is that while $(BE)_g$ is positive (therefore binding) for both HSs and QSs, $(BE)_n$ is always negative for HSs (anti-binding) while it can have both signs for QSs (binding or anti-binding), as shown in tables I-II of Ref.~\cite{Jiang:2019pzw}. To better understand the relation between the sign of $(BE)_n$ and the meaning of self-bound star one should notice that
$(BE)_n$ is always positive definite for QSs of small mass  or in the case quark matter is an incompressible fluid. Since quark matter is compressible, the pressure is not constant but changes with the radius (the local value of the pressure depends both on gravity and on the incompressiblity of quark matter) and $(BE)_n$ can become negative for stars close to the maximum-mass configuration as shown in Ref.\cite{Jiang:2019pzw}. In all cases, if gravity is switched off, the star expands to reach the condition of vanishing pressure. If the thin outer crust of Ref. \cite{Haensel:1986qb} exists, the crust evaporates and the bulk of the star remains bound. If the crust made of strangelets of Ref.\cite{Jaikumar:2005ne} exists, the entire star evaporates into a gas of strangelets and electrons. Finally, if the mixed-phase crust of Ref.\cite{Drago:2001nq} exists the entire object remains bound.

It is remarkable that 
when looking for a universal relation which holds true also for QSs, one has to correlate only $(BE)_g$ to $c$ and not the 
total binding energy $BE$, see Ref. \cite{Jiang:2019pzw} also for other interesting universal relations.

The implications of the previous discussion on
the conversion of HSs into QSs follow immediately:
in general, this process can lead either to smaller objects (having a larger value of $(BE)_g$) or to larger objects (having a smaller value of $(BE)_g$)
as long as the sum of the two binding energies $(BE)^Q_g+(BE)^Q_n$ is larger the the corresponding quantity 
for the progenitor HS. The radius by itself does not play any fundamental role for the exothermicity of the process. For the typical quark matter EoSs adopted for modelling QSs,
the effect of a reduction of the gravitational binding energy (associated with the increase of the radius) does not hinder the conversion, namely the process is still exothermic. This point was already noted in Ref. \cite{Bombaci:2000cv} where, in some cases, it has been found that 
the "negative gravitational conversion energy" was overcome by the nuclear binding energy, thus still allowing for an exothermic conversion process.

These findings in most cases do not depend on the
possible presence of a crust at the surface of the
QS. For instance, the amount of matter in the thin
"outer-crust" of Refs. \cite{Haensel:1986qb,Haensel:2007yy}) is a tiny fraction of the total mass, of the order of $10^{-5} M_\odot$ and its thickness could in
principle add up to a few hundred meters to the total radius of the QS without significantly modifying the energy balance. In the case of the strangelets' composed crust of
Ref. \cite{Jaikumar:2005ne} the thickness is at most of a few ten meters and its contribution to the total mass
and therefore to the energy budget is rather
small. In both those cases the structure of the bulk of
the QS is independent from the crust. In particular, in
the case of the strangelets' composed crust its existence
depends on the value of the surface tension of
quark matter which on the other hand plays no role in
the structure of the bulk of the QS. If the surface
tension is low enough to make the crust energetically
convenient the radius of the star will slightly increase
while its mass will not change significantly. The
hadron-quark mixed phase crust of Ref.\cite{Drago:2001nq} can instead be more relevant for the energy
balance of the conversion since it can contain
a larger fraction of the mass of the star.
While a detailed study on this possibility
could be interesting, it is worth mentioning
that the need of explaining the existence of
very massive QSs strongly constrains the
parameters of the quark matter EoS and it is not clear if the strangelets' crust or the mixed phase crust are still viable possibilities.

\begin{figure}[!ht]
	\begin{centering}
		\epsfig{file=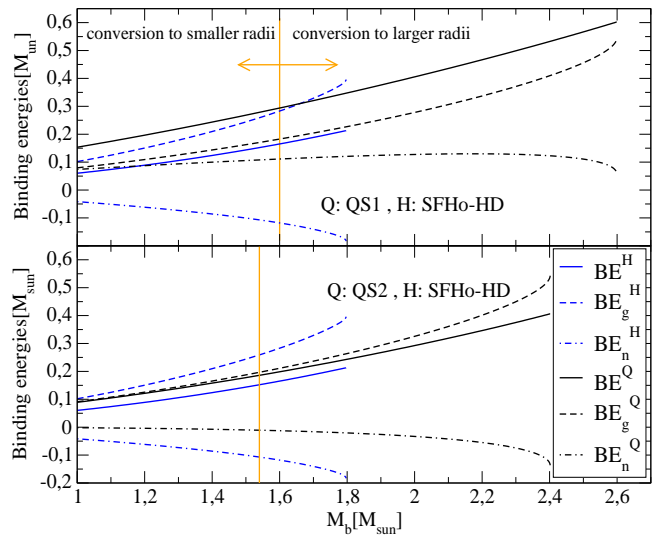,height=8.5cm,width=7cm,angle=-90}
		\caption{Binding energies of QSs and HSs. In the  upper panel QS1 is compared to SFHo-HD and in the lower panel QS2 is compared with SFHo-HD. QSs are always energetically favored with respect to HSs due to the dominance of $(BE)_n$ in the total binding energy.}
	\end{centering}
\end{figure}

Let us discuss some numerical examples. In the following for the hadronic matter we will consider the SFHo-HD EoS of Refs. \cite{Drago:2014oja,Burgio:2018yix,DePietri:2019khb} whose stellar configurations contain hyperons and deltas and we will also consider the SFHo EoS \cite{Steiner:2012rk} for which only nucleons are present. For QSs EoS we use the simplest possibility i.e. a constant speed of sound EoS characterised by the parameters $c_s$, $e_0$ and $n_0$ which correspond to the speed of sound (here set to $\sqrt{1/3}$), the energy density and the baryon density at vanishing pressure, respectively \cite{Drago:2019tbs}.
We call QS1 the set with $e_0/n_0=860$ MeV and $n_0=0.27$fm$^{-3}$ and QS2 the set with $e_0/n_0=930$ MeV and $n_0=0.25$fm$^{-3}$. Both those quark matter EoSs allow to reach the $2M_{\odot}$ limit but the first one corresponds to the case of a large nuclear binding 
whereas the second one to a small nuclear binding (actually it corresponds to the minimum binding needed in order to still have absolutely stable quark matter). 
Let us discuss first 
the behaviour of $BE$ and of its two components $(BE)_n$ and $(BE)_g$ as functions of $M_b$. We display the results in Fig.1 for QS1 and QS2, compared with those for SFHo-HD.
First: the solid black line ($BE^Q$) in both panels lies above the solid blue line ($BE^H)$. The process of conversion is in all cases exothermic no matter whether the QS has a larger or a smaller radius with respect to its progenitor (see the orange vertical lines separating these two possibilities).
Second: $(BE)_g$ (dashed lines) is in all cases positive (binding) and the $(BE)_g$ of HSs is larger than the one of QSs at the same $M_b$. Clearly, 
if only the gravitational binding would be involved in the conversion, the process could not take place because it would be endothermic (even for QSs with small radii). Notice also that for this choice of the hadronic EoS, SFHo-HD, we obtain very compact hadronic configurations i.e. stars with a large value of $(BE)_g$.
The role played by $(BE)_n$ is crucial. Indeed,
$(BE)^H_n$ is large and negative (anti-binding) whereas 
$(BE)^Q_n$ is either positive (thus increasing the binding) for QS1 or negative but very small for QS2 (at least within the range of values of $M_b$ populated by HSs, which is the relevant one when discussing the conversion). Thus,
the strong interactions outweigh the gravitational interactions
in determining the energy balance, as noticed also in \cite{Bombaci:2000cv}.  
This is the crucial requirement for obtaining QSs with larger radii.

\subsection{Conditions for obtaining larger radii: a graphical interpretation}

To study the main issue of this paper, i.e. why the radius of a QS can be larger than the one of the HS having the same baryon mass with a process of conversion that still is exothermic, it is useful to display on the same diagram both the $M_g-R$ and the $M_b-R$ relations. In Fig.2 we show as an example the case of QS1 and SFHo-HD. The two $M_b-R$ relations intersect at the point C: the HS for which $M_b = M_C$ would convert into a QS with exactly the same radius (assuming, as it happens in this case, that the process is energetically favored).

\begin{figure}[!ht]
	\begin{centering}
		\epsfig{file=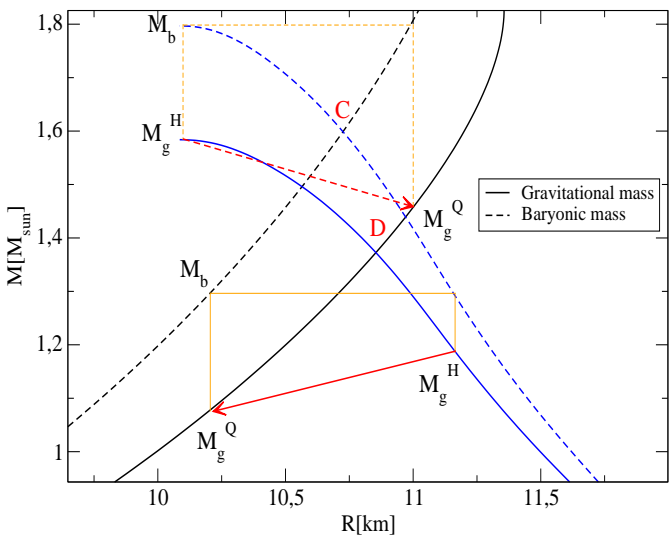,height=8.8cm,width=7cm,angle=-90}
		\caption{Geometrical construction for determining the conversion of a HS into a QS. Solid (black and blue) and dashed (black and blue) lines refer to the $M_g-R$ relations and to the $M_b-R$ relations for QSs and HSs, respectively. The intersection point C between the two $M_b-R$ relations allows to determine if the conversion will lead to a smaller or a larger star. For values of $M_b$ below C (solid orange and red lines) the final configuration is smaller. Conversely, for values of $M_b$ above C the final configuration is larger. Finally, for $M_b$ sitting  exactly on C the two stars have the same radius. Notice that the process is in all three cases exothermic since $M_g^Q < M_g^H$ and that $R_D>R_C$ which is a sufficient condition for obtaining the conversion of HSs to larger QSs.
		The equations of state used for this plot are QS1 and SFHo-HD.}
	\end{centering}
\end{figure}

If one considers values of $M_b$ smaller than $M_C$ than the conversion leads to a QS smaller than the progenitor (the case usually considered in the literature). If instead $M_b$ is larger than $M_C$, the final QS configuration is larger than its progenitor: these two possible transitions are displayed in Fig. 2 and they are indicated by the red solid and dashed arrows, respectively while the orange (solid and dashed) lines indicate the construction of fixing the value of $M_b$ and pinning the corresponding values of $M_g^Q$ and  $M_g^H$. 

Notice that in this example in both the discussed cases the process is exothermic because $M_g^Q<M_g^H$, for two stars having a same $M_b$ and for values of $M_b$ up to the end point of the stability curve of the hadronic configuration. There can be situations in which the conversion is exothermic only up to a maximum value of the baryonic mass, as discussed in the following.

In the above discussion we have indicated how to determine if the QS configuration has a larger or a smaller radius with respect to the hadronic configuration with the same baryonic mass. We have though not yet discussed if there is a simple criterion allowing to determine if conversion process is exothermic, in particular for the case in which the QS would have a radius larger than that of the HS. A sufficient condition can easily be established. Let us call D the point of intersection of the $M_g-R$ curves of the HSs and of the QSs. We will now prove that $R_D > R_C$ is a sufficient condition for having a range of baryonic masses for which an exothermic process produces QSs with radii larger than the corresponding HSs. Indeed if $R_D > R_C$ then
$M_g^H(R_C)=M_g^Q(R_C)+\Delta M$
where $\Delta M >0$.  This last relation can equivalently be expressed as $M_g^H(M_b(R_C))=M_g^Q(M_b(R_C))+\Delta M$. Therefore it exists a value $\epsilon$ such that $M_g^H(M_b(R_C)+\epsilon)>M_g^Q(M_b(R_C)+\epsilon)$. 
For values of $M_b$ in 
the interval $[M_b(R_C),M_b(R_C)+\epsilon]$ one finds
that $R^H(M_b)<R^Q(M_b)$ and $M_g^H(M_b)>M_g^Q(M_b)$. 
Thus the conversion is exothermic even if the radius 
of the QS is larger. On the other hand if $R_C<R_D$ for a similar argument and at least close to $M_b(R_C)$, the conversion to a QS with a larger radius is forbidden because it would be an endothermic process. Only for configurations with significantly larger baryonic mass, $M_b >>M_b(R_C)$, the conversion could in principle take place, depending on the details of the adopted EoSs, but we are not aware of a concrete example of such a situation. Also, from a phenomenological point of view, those cases (large-radius QSs produced from HSs having a mass close to $2 M_{\odot}$) are less relevant, because there is at the moment no suggestion from observations supporting such a scenario.

In conclusion, we found two conditions for having an exothermic transition from a HS to a QS with a larger radius: a necessary condition is that $M_b>M_C$ (in this way the radius is larger) and a sufficient condition is that $R_C<R_D$ (in this way the process is exothermic, at least for masses not much larger than $M_c$).

\subsection{Role of the EoSs: explicit examples}
Let us discuss now the dependence of our results on the choice of the hadronic and quark EoS.
In Fig.3 we present the four combinations of the EoSs discussed above.
We remark that for both hadronic EoSs, the final QS configuration is in principle smaller or larger than its progenitor depending whether $M_b$ is below or above $M_C$. Of course, if $M_b>M_C$ for the transition to be exothermic it must also be $R_D>R_C$, instead the transition is not exothermic at least for $M_b$ close to $M_C$, as discussed above.
From the point of view of the phenomenology, the numerical value of $M_C$ plays an important role. For the SFHo-HD case, $M_{C} \sim 1.6 M_{\odot}$ (at least for the two examples of quark EoSs here considered) and the corresponding gravitational mass of the HS is $\sim 1.4 M_{\odot}$ thus a value sitting at the center of the range of masses for compact stars.
For this choice of a hadronic equation of state, the possibility of generating QSs with large radii through the burning of more compact HSs is therefore plausible and phenomenologically relevant. Conversely,
in the case of SFHo EoS, $M_{C } \sim 2.2 M_{\odot}$ and the gravitational mass of the corresponding HS is $\sim (1.9-2.0) M_{\odot}$, very close to the maximum mass configuration. In that case, the conversion process would lead in most of the cases to a QS having a small radius. The two cases here discussed are presented in the upper (SFHo) and in the lower (SFHo-HD) panel of Fig.3. 
The first case is the one realised in the two-families scenario
in which HSs are very compact and light because of the formation of hyperons and deltas and QSs are massive and with a larger radius. The second case in which HSs are larger than QSs is also possible (with the proper choice of equation of state, see e.g. \cite{Bhattacharyya:2017mdh} for a recent example) but its phenomenology is completely different as we comment in the Conclusions.

\subsection{Conditions for the nucleation of a first droplet of quark matter}
In this paper we have discussed only the macroscopic conditions on energy, i.e. the reduction of the total energy budget, which is necessary to satisfy in order to have an exothermic conversion process of hadronic matter into a quark matter. We have not addressed the trigger for the phase conversion. In the most simple picture, it occurs via nucleation of a seed of strange quark matter into hadronic matter with a significantly large fraction of hyperons. A possible
way to quantitatively define a threshold is to request that the average
distance of strange quarks inside the confined phase is of the
order of, or smaller, than the average distance of nucleons in
nuclear matter \cite{DePietri:2019khb}. The reason is that strange
quarks should be close enough to be able to mutually interact in
order to help the nucleation of a first drop of deconfined quark
matter. Note that this threshold density for deconfinement is
significantly larger than the threshold density for the formation
of hyperons and, from this
viewpoint, this condition agrees with the analysis of e.g. Refs. \cite{Bombaci:2004mt,2009:Bombacijt}, indicating that the first droplet of
quark matter is nucleated at densities larger than the threshold
density of hyperon formation. 
When a sufficiently large strangeness fraction is present at the center of the HS,
for the nucleation process to start it is necessary that the chemical potential of the new phase is smaller than the chemical potential of the old phase, at a same fixed pressure.
This condition is not always realised and actually, as discussed in Ref.\cite{Drago:2019tbs}, it requires EoSs of strange quark matter which are strongly bound, namely the ratio $e_0/n_0$ must be smaller than about $900$ MeV (see also the discussion in Ref.\cite{Bhattacharyya:2017mdh} for the EoS SS2).
Of the two cases here presented, QS2 does not fulfill all the requirements because it is not strongly bound. QS1, having a large nuclear binding, is instead a possible consistent choice for the quark matter EoS. Similarly, for what concerns the hadronic matter EoS, SFHo cannot be considered as a consistent EoS for strange quark matter nucleation, since in SFHo hyperons are not included. There is of course no problem in using SFHo as the hadronic EoS if nucleation of non-strange quark matter is instead considered.

\vskip 0.4cm

\begin{figure}[!ht]
	\begin{centering}
		\epsfig{file=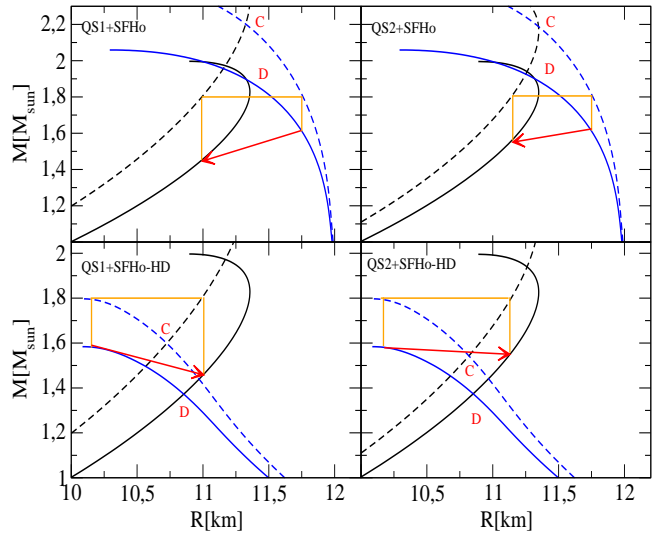,height=8.5cm,width=7cm,angle=-90}
		\caption{$M_g-R$ and $M_b-R$ relations for the four combinations of equations of state are indicated with solid lines and dashed lines respectively. Black lines refer to QSs and blue lines refer to HSs.
		In the upper panel the conversion leads to QSs with radii smaller than their progenitor, in the lower panel, the use of a rather soft hadronic equation of state (which includes hyperons and deltas) leads to QSs which are larger than their progenitors. Upper left panel: $R_C=11.33$ km, $R_D=11.33$ km. Upper right panel: $R_C=11.35$ km, $R_D=11.32$ km. Lower left panel: $R_C=10.73$ km, $R_D=10.86$ km. Lower right panel: $R_C=10.82$ km, $R_D=10.85$ km. }
	\end{centering}
\end{figure}

\section{Conclusions}

We have discussed under which conditions the formation of quark matter inside a compact star can produce an object having a radius larger than its progenitor. We have shown that this is impossible if the final object is a hybrid star, i.e. a star with a crust, while it is possible if a QS is produced, because in the latter case the nuclear interaction can be strongly attractive overcompensating the reduction of the gravitational binding energy. 
Two practical conditions have been found to understand if it is possible to form a QS with a radius larger than that of the progenitor HS: first, if 
the $M_b-R$ relations of the two branches of stars cross at the point C corresponding to a baryonic mass $M_{C}$, it is easily seen that if
the conversion involve stars with $M_b>M_{C}$ the radius of the QS configuration is larger than the one of its progenitor. The opposite is realised if $M_b<M_{C}$. Both possibilities are in principle viable. To decide if the generation of a QS with a larger radius is energetically favored we consider point D, defined as the intersection 
of the $M_g-R$ relations: 
we have shown that if $R_C<R_D$ the conversion to a large QS is exothermic at least for a limited range of masses above $M_C$.

Let us conclude by comparing the scenarios containing a strong phase transition among themselves and with the data. 
The different possibilities can be classified according to two criteria: the content of the second branch which can be a hybrid star or a QS and the mechanism triggering the formation of the new object from the progenitor HS. In the literature two mechanisms have been discussed: one is based on the formation of a droplet of "new phase" at the center of the HS as discussed in Sec.2.4 and on the subsequent expansion of that droplet until the burning of hadrons into quarks is completed \cite{Drago:2015fpa}; the other is based on the rapid increase in the density at the center of a HS or of a hybrid star. To better understand this second possibility, let us refer to the classification of Ref.~ \cite{Alford:2013aca} and consider cases B and D in that paper. If the HS (or the hybrid star) has a mass close to the maximum mass of the first branch (the one mainly composed of hadrons), the further increase of the central density due e.g. to mass accretion or to the slow-down of a rapid rotating object leads to a collapse and to the formation of quark matter which continues till when the amount of quark matter formed is sufficient to stabilize again the star, see Ref.\cite{Mishustin:2002xe}. This possibility has been investigated to explain Supernovae \cite{Hempel:2015vlg} and Gamma-Ray Bursts \cite{Mishustin:2002xe} and has also been recently discussed in the context of mergers \cite{Weih:2019xvw}. In this scenario the formation of quark matter is therefore not associated with the occurrence of a metastable phase (as in the previous case) but to a mechanical instability. In the studies in which this scenario has been investigated it is assumed that the new phase is produced instantaneously and directly in mechanical, thermal and chemical equilibrium with the old phase.
In Table I we summarize these possibilities and we indicate crucial
references where they have been introduced and discussed.
\begin{table}
    \centering
    \begin{tabular}{|c|c|c|}
    \hline
         &QS&hybrid star\\
         \hline
         drop 
         &two-families& hybrid stars\\nucleation
         &-- smaller QSs \cite{Bhattacharyya:2017mdh} &
         by nucleation \cite{Berezhiani:2002ks,Bombaci:2004mt} \\
         &-- larger QSs 
         \cite{Drago:2013fsa}&\\
         \hline
         collapse&not explored&third family
         \cite{Schertler:2000xq,Benic:2014jia}\\
         \hline
    \end{tabular}
    \caption{The different scenarios for the formation of hybrid stars and QSs. }
    \label{tab:my_label}
\end{table}

In the table one the entries is empty, the one corresponding to the formation of QSs via the collapse of the HS. Indeed that possibility has never been discussed and it is not obvious that the system would not directly collapse to a Black Hole, but anyway no analysis has ever been performed on this possibility.
\vskip 0.5cm
\begin{figure}[!ht]
	\begin{centering}
		\epsfig{file=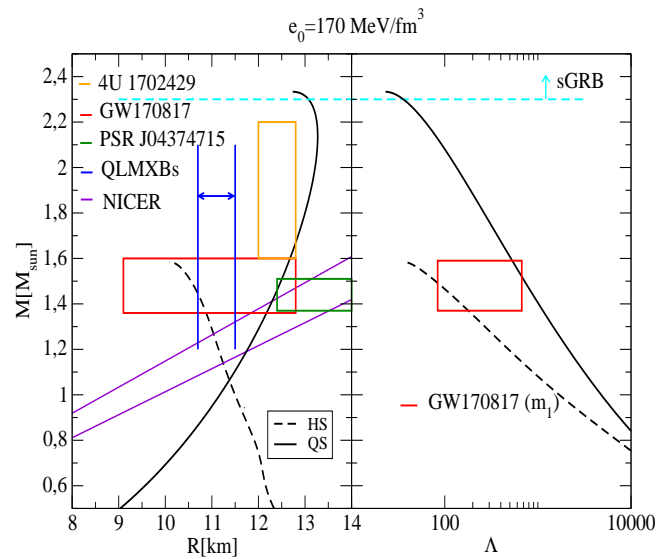,height=8.5cm,width=7.3cm,angle=-90}
    		\caption{
    		Right panel: mass-radius curve of QSs and HSs (as obtained by using the EoS SFHo-HD) and observational constraints on the maximum mass of compact
stars (as obtained by the analysis on sGRBs \cite{Lasky:2013yaa,Lu:2015rta}) and on the radii (as obtained from the direct modelling of the X-ray spectra of
PSR J04374715 \cite{Lasky:2013yaa} (1 $\sigma$- level), of 4U 1702429 \cite{Nattila:2017wtj}(1 $\sigma$- level), of different LMXBs in quiescence \cite{dEtivaux:2019cnf}(1 $\sigma$- level) and from pulse-profile
modeling of X-ray data of millisecond pulsar PSR J0030+0451 \cite{Riley:2019yda} (16\% and 84\% quantiles in marginal posterior mass)). We also display the
constraints on the radius of the most massive star in the merger event GW170817 \cite{PhysRevLett.121.161101} (90\% confidence interval). Right panel: mass -
tidal deformability relation of QSs and observational constraint obtained from GW170817 \cite{PhysRevLett.121.161101} (90\% confidence interval). Here $e_0$ is set to $170$ MeV/fm$^3$ and it is the only parameter affecting the relations between $M_g$ , $R$ and $\Lambda$ ($n_0$ instead modifies the relations between $M_g$ and $M_b$ ).
    	 }
	\end{centering}
\end{figure}

Concerning the comparison with observational data, the third-family and the two-families with small QSs share a similar interpretation of e.g. the mass-radius data since in both cases the HSs have smaller masses and larger radii than the QS or the hybrid star in the new branch. Notice though that, in the third-family scenario, in order to have hybrid stars with small radii the velocity of sound in quark matter needs to be close to the velocity of light  \cite{Alford:2015dpa}, while this request is not necessary in order to obtain QSs with small radii. 
These two schemes are obviously orthogonal to the two-families scenario with large QSs for what concerns the M-R relation, the behaviour of the tidal deformability and of the moment of inertia: these differences open the possibility to distinguish through observations the last scheme from the previous two. For instance the recent paper of Ref.\cite{Christian:2019qer} puts limits on the third family scenario, derived from the analysis of NICER observations \cite{Riley:2019yda,Raaijmakers:2019qny,Miller:2019cac}. Notice that the main conclusion reached in \cite{Christian:2019qer} is that a strong phase transition can take place only if the hadronic EOS is not too soft. This result is correct within the third family scenario, but it does not apply e.g. to the two-families scenario in which the hadronic EOS is very soft (because of the production of resonances) and still a strong phase transition can take place (satisfying both NICER and GW170817 \cite{Char:2019wvo}). In Fig. 4 we show a collection of recent data on masses and radii and how they could be interpreted within the two-families scenario: notice in particular that the data from NICER could correspond either to a HS with a mass of about $1.2 M_\odot$ and a radius of about $11$ km or to a QS with a mass of about $1.4 M_\odot$ and a radius of about $12.5$ km, while within the third-family scenario (or the two-families with small QSs) the object studied by NICER would necessarily be a HS with a not too small radius. While the data discussed in the Figure are in many cases affected by systematics and even the statistical error is not small it is encouraging to notice how a combined analysis can, at least in principle, provide sufficient information to tightly bound the parameters of a model and therefore to provide the possibility to confirm or to invalidate the various interpretative schemes. It is also clear that if eXTP \cite{Zhang:2018edu} will be able to measure masses and radii of a few objects with a precision of the order of $5\%$ the much needed jump in the quality of the data will finally be achieved.

A second and very direct way to discriminate among the various schemes of Tab.1 is to analyze what happens during a merger \cite{Most:2018eaw,Blacker:2020nlq}. In the third-family scenario the production of quarks takes place without any delay once a critical density is reached and therefore in a merger the hypermassive configuration, produced by differentially rotating the quark-rich object, can always be generated unless the total mass of the merger exceeds the mass of the hypermassive configuration. Also, the quark-rich object is more compact than the quark-poor one and therefore, after that density threshold is reached, the GW signal has a higher frequency than the one in the absence of quark deconfinement \cite{Bauswein:2018bma,Weih:2019xvw}. 
The two-families scenario with larger QSs behaves in a completely different way. First, the production of quarks is not instantaneous and the combustion of the central region of the star from hadrons into quarks takes a few milliseconds \cite{Drago:2005yj,Herzog:2011sn,Pagliara:2013tza,Drago:2015fpa}. Therefore if the collapse to black hole takes place in less than a few milliseconds quarks cannot be produced and in that situation the mass of the hypermassive configuration is based on the HS and not on the QS and it is therefore significantly smaller than in the cases in which the maximum mass of the HS can reach 2$M_\odot$. This implies that in the case of a HS-HS merger the post-merger object can directly collapse to a black-hole even for total masses smaller than the one of GW170817 (that in the two-families scenario was the merger of a HS with a QS) \cite{Drago:2017bnf,DePietri:2019khb}. Notice that the two-families scenario with smaller QSs \cite{Bhattacharyya:2017mdh} does not predict this possibility because, even though quark deconfinement is not instantaneous also there, if quarks are not produced the newly formed object made of hadrons is not soft and its maximum non-rotating mass  can reach or exceed 2 $M_\odot$ (and the same argument applies to the case in which hybrid stars are produced by quark nucleation). Finally, only in the two-families scenario with larger QSs the GW signal has a lower frequency once quarks have deconfined \cite{Bauswein:2015vxa,DePietri:2019khb}. These two features of the two-families scenario with large radii of the QSs are therefore characterizing that scheme from all the others and they can be tested by a combined GW and EM analysis of future events.


\end{document}